\documentclass[a4paper]{jpconf}
\usepackage{bm}
\usepackage{mathrsfs}
\usepackage[all]{xy}
\newcommand{\be}{\begin{equation}}
\newcommand{\ee}{\end{equation}}
\newcommand{\bea}{\begin{eqnarray}}
\newcommand{\nn}{\nonumber}
\newcommand{\eea}{\end{eqnarray}}

\begin{document}

\title{Extending Sibgatullin's ansatz for the Ernst potential
 to generate a richer family of axially
symmetric solutions of Einstein's equations}

\author{T P Sotiriou\footnote[1]{Present address: SISSA, International School for
Advanced Studies, Via Beirut, 2-4 34014 Trieste, Italy} and G Pappas}
\address{Section of Astrophysics, Astronomy, and Mechanics,\\
Department of Physics, National and Kapodistrian University of Athens,\\
Panepistimiopolis, Zografos GR-15783, Athens, Greece.}

\ead{tsotiri@phys.uoa.gr; sotiriou@sissa.it; gpappas@phys.uoa.gr}

\begin{abstract}

The scope of this talk is to present some preliminary results on
an effort, currently in progress, to generate an exact solution of
Einstein's equation, suitable for describing spacetime around a
rotating compact object. Specifically, the form of the
Ernst potential on the symmetry axis and its connection with the
multipole moments is discussed thoroughly. The way to calculate
the multipole moments of spacetime
directly
from the value of the
Ernst potential on the symmetry axis is presented. Finally, a mixed
ansatz is formed for the Ernst potential including parameters
additional to the ones dictated by Sibgatullin. Thus, we 
believe
that this talk can also serve as a comment on choosing the
appropriate ansatz for the Ernst potential.
\end{abstract}

\section{Introduction}

In order to analytically describe spacetime around a rotating
compact object in the framework of general relativity, one needs an
exact asymptotically flat solution of Einstein-Maxwell equations.
To simplify the problem it is reasonable to assume that a
long-lived, quiescent rotating body 
possesses axial symmetry
and
reflectional symmetry with respect to the equatorial plane. 
The solution must 
depend on
a number of arbitrary physical parameters
such as the mass, the angular momentum, the mass-quadrupole
moment, the magnetic dipole e.t.c.
of the central body.

Sibgatullin has proposed a way to generate stationary axisymmetric solutions
\cite{sib1}, starting from the value of the Ernst potentials
\cite{ernst1,ernst2} $\mathcal{E}$ and $\Phi$ on the symmetry
axis. His method has been used successfully by many authors (see
for example \cite{sib2,manko}). However the arbitrary parameters
appearing in the proposed ansatz for $\mathcal{E}$ and $\Phi$ are
physically meaningful only in specific and rather special cases
(for example the Kerr solution).

On the other hand, it is well known for quite some time that
spacetime geometry around an axisymmetric object leads to a
specific set of the Geroch-Hansen multipole moments \cite{geroch,
hansen}. Thus, the scalar moments can be used to characterize the
spacetime, in the same way that the newtonian gravitational
moments characterize the newtonian gravitational field. In this
paper we will try to combine 
multipole moments with
Sibgatullin's arbitrary parameters in order to form an ansatz for
the Ernst potential $\mathcal{E}$ on the symmetry axis, suitable
to obtain a metric for an axisymmetric compact object through
Sibgatullin's method. Due to lack of space and time this study
will be restricted to vacuum spacetime. Note however that it can
be easily generalized to include electrovacuum spacetimes. 

The plan of this paper is the following: In Sec. \ref{sibmeth} we give an outline of Sibgatullin's
method. In Sec. \ref{moments} we show the way to calculate the moments  
directly 
from the value of the
Ernst potential on the symmetry axis. We also present a way to generate a solution with specific
multipole moments. In Sec. \ref{mixed} we present a ``mixed'' ansatz for the Ernst potential
including both Sibgatullin arbitrary parameter and a number of new parameters related to the moments.
Finally we comment on the restrictions that come up if we additionally impose reflectional symmetry with
respect to the equatorial plane.

\section{Sibgatullin's method for generating exact solutions}
\label{sibmeth}

It is known for quite some time that in the case of a stationary, axisymetric spacetime the 
metric functions can be fully determined if the so called Ernst potential $\mathcal{E}$ is known
\cite{ernst1}. Sibgatullin has proposed a way of generating solutions \cite{sib1}
of Einstein's field equations, that reduces to evaluating the
solution of a linear system of algebraic equations. His approach
is based on the definition of the Ernst potential on the symmetry
axis in terms of rational functions, involving a series of
parameters $a_j $ \cite{sib2}

\be \label{eq1} \mathcal{E}(\rho = 0) = e(z) = \frac{z^n - Mz^{n - 1} +
\sum\limits_{j = 1}^n {a_j z^{n - j}} }{z^n + Mz^{n - 1} +
\sum\limits_{j = 1}^n {a_j z^{n - j}} } \ee

We will merely outline some key points of the method. Sibgatullin
has shown that the Ernst potential $\mathcal{E}(\rho ,z)$ can be evaluated
from the potential (\ref{eq1}) 
which is given 
on the symmetry axis, as the
integral 
\be \mathcal{E}(\rho ,z) = \int\limits_{ - 1}^1 {\frac{\mu (\sigma
)e(\xi )d\sigma }{\sqrt {1 - \sigma ^2} }} \ee
 were the function
$\mu (\sigma )$ satisfies the following linear integral equation and normalization condition
respectively:
\be {\rm P}\int\limits_{ - 1}^1 {\frac{\mu (\sigma )\left( {e(\xi ) +
\tilde {e}(\xi )} \right)}{(\sigma - \tau )\sqrt {1 - \sigma ^2}
}d\sigma } = 0,\qquad \int\limits_{ - 1}^1
{\frac{\mu (\sigma )d\sigma }{\sqrt {1 - \sigma ^2} }} = \pi \ee
were the P denotes the principal value of the integral, $\xi
\equiv z + i\sigma \rho ,\quad
\sigma ,\tau \in \left[ { - 1,1} \right]$ and the functions $e(\xi
)$ and $\tilde {e}(\xi )$ are analytical continuations on the
complex plane for the functions $e(z)$ and $e^\ast (z)$. The first
step is to express equation (\ref{eq1}) as an expansion of
elementary fractions $e(z) = 1 + \sum_{j = 1}^n {\frac{e_j
}{z - \tilde {a}_j }} $ and the sought for solution of the integral
equation as $\mu (\sigma ) = A_0 + \sum_{k = 1}^{2n}
{\frac{A_k }{\xi - \xi _k }} $ were the $\tilde {a}_j $ and $\xi
_k $ are the roots of the polynomials $z^n + Mz^{n - 1} +
\sum_{j = 1}^n {a_j z^{n - j}} = 0$ and 
$e(z)e^\ast(z)=0$
 respectivly
. Substituting 
the series expansions of 
$e(z)$ and $\mu (\sigma )$ in the integral
equation and the normalization condition we obtain a closed system
of $2n + 1$ linear algebraic equations for the $2n + 1$ parameters
$A_k $.
 These coefficients
can be evaluated and they completely determine the Ernst potential
that will generate the solution. In particular, it has been shown
by Ruiz, Manko and Martin \cite{manko2} that the metric functions 
can be evaluated with the use of some
determinants which are functions of the parameters $e_j, a_j, \xi
_k$.

\section{The multipole moments of the solutions}
\label{moments}

In \cite{fodor} an algorithm is presented for the evaluation of the multipole moments of 
stationary axisymmetric spacetimes from the series expasion of the function $\tilde{\xi}$ on the symmetry 
axis. The real and the imaginary part of $\tilde{\xi}$ are related to the so-called Hansen potentials
respectively \cite{hansen}. 
Since the multipole moments are evaluated at infinity, we will use the prefered coordinates
\begin{equation}
\label{eq18}
\bar {\rho } = \frac{\rho }{\rho ^2 + z^2},\qquad\bar {z} = \frac{z}{\rho ^2 +
z^2}.
\end{equation}
where $\rho$ and $z$ are Weyl coordinates. As mentioned in \cite{fodor} $\tilde{\xi}$ can expanded
around $\bar{z}=0$ in the following way.
\begin{equation}
\label{eq21}
\tilde {\xi }(\bar {\rho } = 0) = \sum\limits_{n = 0}^\infty {m_n \bar
{z}^n}
\end{equation}
The multipole moments can then be evaluated from the $m_{n}$'s using the relations 
presented in p. 2255 of \cite{fodor}.

If we express $\tilde{\xi}$ in terms of the Ernst potential $\mathcal{E}$ then
\be
\label{comb1}
\tilde{\xi}=\sqrt{\rho^2+z^2}\frac{1-\mathcal{E}}{1+\mathcal{E}},\quad\tilde{\xi}(\rho=0)=z\frac{1-\mathcal{E}(\rho=0)}{1+\mathcal{E}(\rho=0)},
\ee
or in the prefered coordinates
\be
\label{comb4}
\tilde{\xi}(\rho=0)=\bar{z}^{-1}\frac{1-\mathcal{E}(\rho=0)}{1+\mathcal{E}(\rho=0)}.
\ee
It is now obvious that the moments can be evaluated 
 from $\mathcal{E}(\rho=0)$.
Using (\ref{eq21}) and (\ref{eq1}) we get
\be
\label{comb5b}
\sum\limits_{i = 0}^\infty {m_i \bar
{z}^i}
=\frac{M}{1+\sum\limits^{n}_{j=1}{a_{j}\bar{z}^{j}}}.
\ee

Equation (\ref{comb5b}) can be used to relate $m_{i}$ with $a_{j}$ and vise versa. Applying Taylor expansion 
at $\bar{z}=0$  
we derive the following recursive formulae:
\bea
\label{comb8}
m_{0}=M,\quad m_{n}=-\sum\limits_{l=1}^{n}{a_{l}m_{n-l}}\quad\textrm{or}\quad a_{0}=1,\quad a_{n}=-\sum\limits_{l=1}^{n}{\frac{m_{l}}{M}a_{n-l}}.
\eea
To derive the second 
set of expressions
, of course, one has to regard that the number of $m_n$'s is
finite and the number of $a_{n}$'s is infinite.
Before we make any comments we present two examples. If
we set
\be
\label{comb10}
a_{n}=0,\quad n=1,2,3,\ldots
\ee
then it follows that
\bea
\label{comb11}
m_{0}=M,\quad m_{n}=0,\quad n=1,2,3,\ldots
\eea
which will lead to the Schwarzchild solution, 
while 
if we set
\bea
\label{comb12}
a_{1}=-ia,\quad a_{n}=0,\quad n=2,3,4,\ldots
\eea
then it follows that
\bea
\label{comb13}
M_{2n}=(-1)^{n}Ma^{2n},\quad
J_{2n+1}=(-1)^{n}Ma^{2n+1},\quad n=0,1,2,3,\ldots
\eea
which are the mass and the mass current moments of the Kerr solution respectivelly. 

We have related $\mathcal{E}(\rho=0)$ and thus Sibgatullin's arbitrary parameters with the moments. 
This is another way to demonstrate that all information regarding spacetime are included in 
the value of the Ernst potential on the symmetry axis. Using eqs. (\ref{comb8}) one can 
compute the moments without much effort even before generating the solution. One can also 
get an insight for the number and the nature of the $a_j$'s needed to generate a solution with 
a certain set of multipoles. However, it is important to say that 
a physical object is in principle expected to have 
an infinite number of moments. It is easy to see that a finite nunber of 
$a_{j}$'s 
imposes a relation between the different order multipoles. The more 
$a_{j}$'s 
one uses the more flexible 
this relation becomes. However every new 
$a_{j}$ 
included leads to an increace of the order of the polynomials 
of the Ernst potential, and consequently creates serious difficulties in 
generating the corresponding metric.

Before we close this section it is important to notice that Sibgatullin's method can be also used to
generate an exact solution with a desired set of multipole moments or equivelantly a desired set of 
$m_i$'s. If we solve (\ref{comb4}) for 
$\mathcal{E}(\rho=0)$ we get
\be
\label{sibmom}
\mathcal{E}(\rho=0)=\frac{1-\bar{z}\sum\limits_{i = 0}^\infty {m_i \bar{z}^i}}{1+\bar{z}\sum\limits_{i = 0}^\infty {m_i \bar{z}^i}}=
\frac{1-\sum\limits_{i = 0}^\infty {m_i z^{-(i+1)}}}{1+\sum\limits_{i = 0}^\infty {m_i z^{-(i+1)}}}.
\ee
We can use (\ref{sibmom}) instead of (\ref{eq1}) and follow the procedure described in section \ref{sibmeth}
to obtain the solution. We 
could turn 
the infinite sum of eq. (\ref{eq21}) into a finite one, by
setting all the moments of order $>n$ equal to zero. 
This, even though not 
exact, can most of the times 
be considered as a good 
approximation.

\section{A mixed ansatz for the Ernst Potential}
\label{mixed}

Sibgatullin's ansatz for $\mathcal{E}(\rho = 0)$ is rather general, but not the easiest one 
can use to generate a solution as we will show. Consider the following
ansatz for $\mathcal{E}(\rho = 0)$:
\be \label{mixans} \mathcal{E}(\rho = 0) = \frac{z^n - Mz^{n - 1} +
\sum\limits_{j = 1}^n {a_j z^{n - j}}-\sum\limits_{i = 1}^n {b_i z^{n-1-i}}
 }{z^n + Mz^{n - 1} +
\sum\limits_{j = 1}^n {a_j z^{n - j}}+\sum\limits_{i = 1}^n {b_i z^{n-1-i}} } \ee
It is easy to see that if $a_j=0$ for all $j$'s then $b_i=m_i$ and if all $b_i=0$ the ansatz is 
reduced to Sibgatullin's
one. 
It is obvious that it is not possible to define a new finite set of parameters $a'_{j}$ in order 
to bring the ansatz in the form of eq. (\ref{eq1}). 
This simply depicts the close relation of the $b_i$'s with the $m_i$'s.
Notice that one can use the above ansatz to generate an exact solution with Sibgatullin's method.
No modifications in the generation algorithm are necessary.
The merit of using such a mixed ansatz with $l$ $a$'s and $n$ $b$'s lies in the fact 
that the degree of its polynomials will be equal either to $l$ or to $n+1$ (depending on which one is the 
largest), whereas in the original 
ansatz it is equal to $k=l+n$ in order for it to have the same number of parameters. 
This of course can lead to serious simplification 
of the computations needed to generate the solution. For example a 6 parameter solution would 
require third degree polynomial instead of fifth.

Before we proceed lets take a moment to focus on the special case in which spacetime has an additional
symmetry: reflection symmetry with respect to the equatorial plane. This is a realistic assumption 
for most astrophysical objects. In this case $m_i$ should be real for even and imaginary for 
odd $i$'s \cite{hansen, ryan, sotapo2}. By evaluating the $m_i$'s in terms of the $a_i$'s and $b_i$'s 
one can show that, if there is a finite number of $a_i$'s and $b_i$'s  then they should also 
be real for even $i$'s and imaginary for odd $i$'s.

A simple case for an ansatz is the following:
\be
\label{our}
\mathcal{E}(\rho=0)=\frac{z^2-Mz-iaz+bc-icM}{z^2+Mz-iaz+bc+icM},
\ee
where all parameters are real. The first five multipole moments of the corresponding spacetime 
are the following:
\bea
\label{moments2}
M_0&=&M,\quad M_1=0,\quad M_2=-(a^2+ac+bc)M,\quad M_3=0,\nn\\
 M_4&=&\left( a^4 + a^3 c + 3 a^2 b c + 2 a b c^2 + 
     b^2 c^2 \right) M - 
  \frac{1}{7}\left[M \left( {\left( a + c \right) }^2 M^2 - 
       \left( a^2 + a c + b c \right) M^2 \right) \right]\nn\\
J_0&=&0,\quad J_1=(a+c)M,\quad J_2=0,\nn\\
J_3&=&- a^3 M  - a^2 c M - 2 a b c M - 
  b c^2 M,\quad J_4=0.
\eea
If we set $c=0$ we get
\bea
M_{2n}=(-1)^{n}Ma^{2n},\quad
J_{2n+1}=(-1)^{n}Ma^{2n+1},\quad n=0,1,2,3,\ldots.
\eea
On the other hand if we set $b=a+c$ 
\bea
M_{2n}=(-1)^{n}M(a+c)^{2n},\quad
J_{2n+1}=(-1)^{n}M(a+c)^{2n+1},\quad n=0,1,2,3,\ldots.
\eea
Therefore, there are two discrete ways for the metric to be
reduced to
the Kerr metric, 
having different angular momentum. The first one is by reducing the ansatz to the known ansatz for the Kerr
metric by setting $c=0$ as we showed. The second one which
does 
also reduce to a Kerr metric with angular momentum $c$ if $a=0$, is by setting $b=a+c$. From eq. (\ref{moments2}) one can see that 
by using freely 
the parameters $a$, $b$, $c$ he can give the desired 
value to the angular momentum and the mass quadropole and still have two free parameters including the 
mass. Thus we think that starting from this ansatz it will be possible to generate a solution suitable 
to describe spacetime around compact rotating objects sufficiently 
well.

Finally, it is worth mentioning that this whole approach can be generalized to include electrovacuum 
spacetimes (based on \cite{ernst2, sotapo2, sotapo}).

\ack

The authors would like to thank Theocharis Apostolatos for suggesting this project and for all his
valuable comments. This work was partly supported by the ``PYTHAGORAS'' research funding program
Grant No 70/3/7396.


\section*{References}
\begin{thebibliography}{25}
\bibitem{sib1} Sibgatullin N R 1984 {\it Oscilations and Waves in Strong
Gravitational and Electromagnetic Fields} (Nauka, Moscow, 1984;
English translation: Springer-Verlag, Berlin, 1991)
\bibitem{ernst1} Ernst F J 1968 {\it Phys. Rev.} {\bf 167} 1175
\bibitem{ernst2} Ernst F J 1968 {\it Phys. Rev.} {\bf 168} 1415
\bibitem{sib2} Siibgatullin N R and Sunyaev R A 1998 {\it Astronomy Letters} {\bf 24} 894
\bibitem{manko} Manko V S, Sanabria-G\'omez J D, Manko O V 2000 {\it Phys. Rev.} D {\bf 62} 044048
\bibitem{geroch} Geroch R 1970 {\it J. Math. Phys.} {\bf 11} 2580
\bibitem{hansen} Hansen R O 1974 {\it J. Math. Phys.} {\bf 15} 46
\bibitem{manko2} Ruiz E, Manko V S, Martin J 1995 {\it Phys. Rev.} D {\bf 51} 4192
\bibitem{fodor} Fodor G, Hoenselaers C, Perj\'es Z 1989 {\it J. Math. Phys.} {\bf 30} 2252
\bibitem{ryan} Ryan F D 1995 {\it Phys. Rev.} D {\bf 52} 5707
\bibitem{sotapo2} Sotiriou T P and Apostolatos T A (submitted to {\it Phys. Rev.} D) {\it Preprint:} gr-qc/0410102
\bibitem{sotapo} Sotiriou T P and Apostolatos T A 2004 {\it Class. Quantum Grav.} {\bf 21} 5727
\end{thebibliography}
\end{document}